# Spatial Equity of Micromobility Systems: A Comparison of Shared E-scooters and Station-based Bikeshare in Washington DC

**Lin Su**
Department of Civil & Coastal Engineering
University of Florida
1949 Stadium Rd, Gainesville, FL 32611, United States
Email: sulin@ufl.edu
ORCID: https://orcid.org/0000-0001-5507-0389

**Xiang Yan, Ph.D.**
Department of Civil & Coastal Engineering
University of Florida
1949 Stadium Rd, Gainesville, FL 32611, United States
Email: xiangyan@ufl.edu
ORCID: https://orcid.org/0000-0002-8619-0065

**Xilei Zhao, Ph.D.**
Department of Civil & Coastal Engineering
University of Florida
1949 Stadium Rd, Gainesville, FL 32611, United States
Email: xilei.zhao@essie.ufl.edu
ORCID: https://orcid.org/0000-0002-7903-4806

**Acknowledgment**

This research was supported by the U.S. Department of Transportation through the Southeastern Transportation Research, Innovation, Development and Education (STRIDE) Region 4 University Transportation Center (Grant No. 69A3551747104).

**Author Contribution**
Lin Su: Conceptualization, Methodology, Data curation, Formal analysis, Investigation, Writing- Original draft preparation.
Xiang Yan: Conceptualization, Methodology, Formal analysis, Investigation, Writing- Original draft preparation.
Xilei Zhao: Investigation, Writing- Reviewing and editing, Supervision, Funding acquisition.

**Conflict of Interest**
None declared.

# Spatial Equity of Micromobility Systems: A Comparison of Shared E-scooters and Station-based Bikeshare in Washington DC

**Abstract**
Many cities around the world have introduced dockless micromobility services in recent years and witnessed their rapid growth. Shared dockless e-scooters have the potential to benefit neighborhoods that lack access to station-based bikeshare services, but they may also exacerbate the existing spatial disparities. While some studies have examined the equity of station-based bikeshare systems, limited knowledge is available regarding dockless e-scooter services. This study uses Washington DC as a case study, a city with both dockless e-scooter and station-based bikeshare systems, to conduct equity analysis of the two types of micromobility options. We develop an analytical framework to examine how dockless e-scooter and station-based bikeshare differ on a set of equity-related outcomes (i.e., availability, accessibility, usage, and idle time) across neighborhoods of different socioeconomic categories. Results reveal that dockless e-scooter services increase accessibility to shared micromobility options for disadvantaged neighborhoods but also widen the access gap across neighborhoods. Compared to bikeshare, shared e-scooters have a higher level of spatial accessibility overall due to greater supply; however, the greater supply largely leads to longer average idle time of shared e-scooters rather than a greater number of trips. Finally, it appears that the bikeshare system's equity program effectively promotes low-income use but e-scooters' equity programs do not. Our findings suggest that increasing vehicle supply alone would probably not lead to higher micromobility use in disadvantaged neighborhoods. Instead, policymakers should combine a variety of strategies such as promoting the enrollment of equity programs and reducing access barriers (e.g., smartphone and banking requirements) to micromobility services.

Keywords: Micromobility, dockless e-scooter, bikeshare, transportation equity

# 1. Introduction

In recent years, shared micromobility services have experienced rapid growth in cities across the globe, expanding transportation options for many travelers. However, low-income populations are often excluded from the broad benefits of these services due to a variety of barriers such as lack of spatial access (physical barrier), unaffordability (financial barrier), and without a bank account, a smartphone, or a data plan (logistical barrier), or lack of digital literacy to access micromobility services (technological barrier) (Dill and McNeil, 2021; Kodransky and Lewenstein, 2014). Understanding the equity of micromobility services is of major importance to transportation planners and policymakers.

So far, researchers have mostly focused on the equity of station-based micromobility systems (Brown et al., 2019; Qian and Jaller, 2020; Qian and Jaller, 2021). Previous studies have shown that bikeshare users are disproportionately White, male, and wealthier, and people of color are often underserved (Buck et al., 2013; Fishman, 2016). For instance, a recent survey conducted by the Washington Area Bicycle Association found that bikeshare used by low-income travelers was lower compared to the general population (Kodransky and Lewenstein, 2014). A plausible reason is that low-income individuals lack convenient spatial access to bikeshare due to insufficient station density. Some researchers have analyzed the inequalities in the spatial distribution of mobility across neighborhoods of different socio-economic characteristics. These studies have consistently found that residents of low-income neighborhoods are disproportionately underserved. For instance, Chen et al. developed a modeling approach to measuring individuals' accessibility in southern Tampa and found that accessibility to bikeshare was not evenly distributed across neighborhoods of different socio-demographic statuses (Chen et al., 2019). The authors observed lower accessibility in Black and middle-income neighborhoods. Similarly, one study that focused on the City of Seattle showed greater bike availability in socioeconomically advantaged neighborhoods with higher median incomes and more college graduates (Mooney et al., 2019).

Due to their recent emergence, shared e-scooter services have not been thoroughly studied regarding their equity implications. Free-standing with fewer spatial restrictions, shared dockless e-scooters have the potential to enhance spatial accessibility to destinations and to close the network gap of station-based systems. On the other hand, since shared e-scooters services are usually operated by for-profit private entities, the operators tend to put less emphasis on promoting equity. E-scooter trips tend to be more expensive than bikeshare trips, and so some travelers would be excluded from using e-scooters. Moreover, e-scooter companies may decide to place fewer e-scooters in low-income and minority neighborhoods, resulting in lower availability and accessibility of e-scooter services in these neighborhoods and hence lower e-scooter use. To date, there has been limited empirical evidence that can shed light on these equity-related issues.

Specifically, we address the following research questions: *Does the dockless e-scooter program provide equitable services across neighborhoods? Are shared e-scooter services more equitable compared to the station-based bikeshare program?*

We focus on Washington DC as the study area, where the bikeshare system (Capital Bikeshare) is one of the most popular in the U.S. and thousands of e-scooters have been permitted on its streets since 2017. We analyzed several weeks of data in June 2019 and July 2019, when both micromobility systems were widely deployed across Washington DC neighborhoods and had high utilization rates. The bikeshare system and the e-scooter system had a different pricing mechanism at the time. A single bikeshare trip was charged at 30-min increments (e.g., $2 per 30-min increment), and a variety of pass options (day pass or 30-day pass) and annual membership options (regular, corporate, and equity program) were offered. By contrast, e-scooter trips were charged on a per-min basis (varies between $0.24 to $0.39 per min by operator) plus a $1 unlock fee. The district also required e-scooters operators to provide discounted programs to promote e-scooter use among qualified low-income travelers.

We develop an analytical framework to examine how dockless e-scooter and station-based bikeshare differ regarding a set of equity-related outcomes (i.e., availability, accessibility, usage, and idle time) across neighborhoods in different socio-economic categories, which will be discussed in detail below. We use the block group as the main unit of analysis in this study, and we categorize block groups by Equity Emphasis Area (EEA) status and by income levels and racial compositions according to the most recent census data.

The rest of the paper is structured as follows. The next section introduces the analytical framework, followed by a section that describes the data sources. The following sections summarize the results and discuss the implications. We conclude the study by summarizing key findings, suggesting policy implications, and noting the limitations.

## 2. Analytical Framework

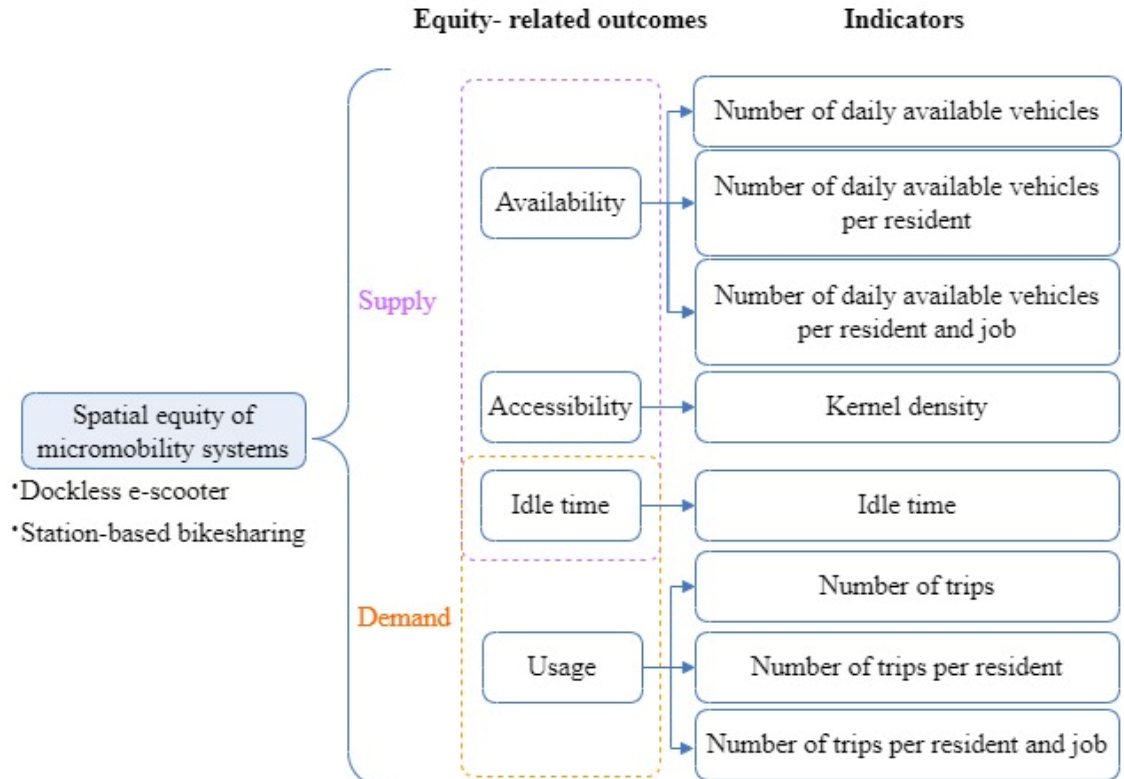

Figure 1. The analytical framework of equity in micromobility services.

Figure 1 illustrates the analytical framework of our study. We examine the following equity-related outcomes. First, we evaluate the supply of micromobility services across space with measures of *availability* and *accessibility*. Availability is defined as the number of available micromobility vehicles in a geographic unit; and in this study, we use census block groups as the geographic unit of analysis. Accessibility quantities the number of micromobility vehicles reachable from a block group, which complements the availability measure by not excluding micromobility vehicles located at a reachable distance from the block group boundary. The two measures have pros and cons, and both have been examined in previous studies (Meng and Brown, 2021). Accessibility is a more theoretically sound supply measure than availability, but availability is easier to interpret, especially for the general public; moreover, availability is better tied to micromobility regulations regarding minimal or maximum fleet size in specific zones. Second, we evaluate the demand aspect of micromobility services, that is, we measure micromobility usage with the number of trips. Third, we further examine *idle time*, which indicates the time duration when a micromobility vehicle is not used, i.e., the time interval between the end timestamp of a trip and the start timestamp of the following trip. Idle time is a measure of operational efficiency, as it is jointly shaped by the supply of and demand for micromobility services. It can also help inform the effectiveness of enhancing supply on increased micromobility use; when low use is coupled with long idle time, policy measures that increase supply can be ineffective.

To further explore the geographic equity of micromobility system, we analyze how the above measures vary across different categories of block groups. We divide block groups by Equity Emphasis Area (EEA) status, by income levels, and by racial and ethnic compositions according to the American Community Survey (ACS) 2014-2019 5-year estimates. Whether a block group belongs to the Equity Emphasis Area is determined by the National Capital Region Transportation Planning Board, which considers if there is a significant concentration of low-income and minority populations including African American, Asian, and Hispanic or Latino (National Capital Regional Transportation Planning Board, 2017). Figure 2 shows the block groups that fall within the EEA. Moreover, we classify block groups by quartile of household median income: block groups with a median household income of $49,222 or below as low income, those with a median household income of between $49,223 and $130,614 as middle income, and those with a median household income of $130,615 or above as high income. Figure 3 shows the distribution of median household income in DC. Finally, we classify block groups into racial and ethnic majority categories based on the ACS data; that is, we define block groups where more than 50% of the residents identify themselves as a single race or ethnicity of X (e.g., White, Black, and Hispanic) as "X-majority" block groups (Brown, 2019). Block groups without a single race or ethnicity representing more than 50% of the population are categorized as "No majority."

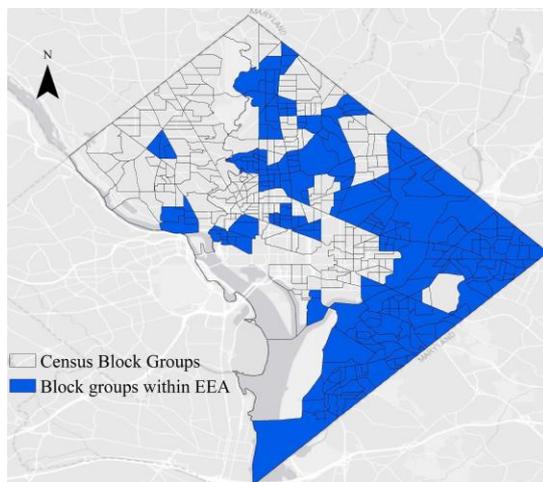 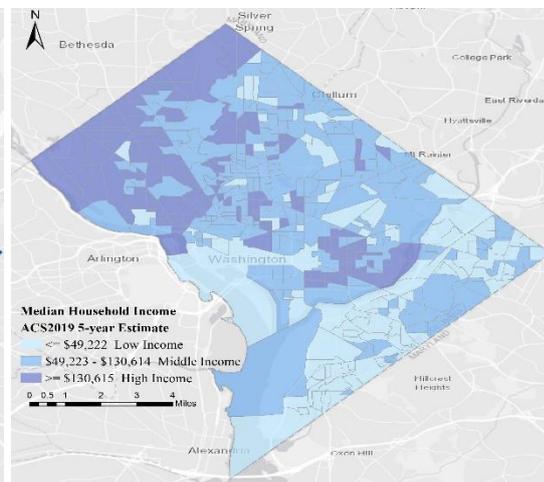

Figure 2. Distribution of EEA in DC.    Figure 3. Median Household Income in DC.

The threshold approach (i.e., classifying neighborhoods into various groups based on some threshold values) that we adopt here to classify disadvantaged/advantaged communities has some limitations. Notably, this approach ignores within-neighborhood variations in population characteristics and hence may not work well when population groups are not spatially segregated. Rowangould et al. (2016) proposed a population-weighted approach (i.e., measuring outcomes for each population group with a weighted mean across all neighborhoods) to address this issue. However, the population-weighted approach is appropriate for measuring some of our equity outcomes (availability and accessibility) but not others (idle time and usage). Hence, we use the threshold approach here to ensure consistency.[1]

An important implication of adopting this threshold approach is that the analysis conducted in this study will mainly shed light on the question of equity across places rather than across population groups. That is, we will examine the disparities in measured outcomes across *block groups*—rather than across *micromobility users*—in various socioeconomic categories. One should not assume that trips generated from block groups in a certain socioeconomic category are made by people belonging to the same socioeconomic category; for instance, it is very likely that many trips happened in low-income block groups are not made by low-income people. Despite this limitation, we argue that the analysis of spatial equity of micromobility systems has many merits. Notably, micromobility regulations are often place-based rather than people-based. For example, Washington DC requires that each operator shall balance its fleet of shared devices by having deployed at least 3% of permitted fleet in each ward between 5 am to 7 am (The department of transportation Washington, D.C., 2021.); in Seattle, the vendors are required to distribute at least 10% of its deployed fleet in designated Equity Focus Neighborhoods (The Seattle Department of Transportation, 2022.).

---

[1] We have also computed availability and accessibility indicators with the population-weighted approach (see Appendix). The results change slightly but the main findings are largely the same.

## 3. Data

The main data sources for this study include the following: e-scooter data, docked bikeshare data, and socio-demographic and employment data.

### 3.1 E-scooter Data

E-scooter data from June and July in 2019 are collected from real-time public APIs provided by each operator, and the obtained data are in General Bikeshare Feed Specification (GBFS) format. The GBFS data indicate the precise location of the available e-scooters or e-bikes at a given moment, based on which we compute the supply-related indicators discussed in Figure 1. Specifically, we compute the e-scooter availability and accessibility measures using the GBFS data scrapped at 6:00 am, which is the start of the day when e-scooter operators start to provide services. We further infer e-scooter trips from the GBFS data for four operators: Spin, Lime, Bird, and Lyft (the available data for other operators such as Helbiz cannot be used to infer trips). Xu et al. (2020) provides a detailed description of the trip inference algorithms adopted in this study (Xu et al., 2020). For the Lime and Spin data, each scooter trip record contains the trip's start date and time, end date and time, trip duration, trip distance, start and end location, and vehicle ID. The data has been processed to remove trips lasting less than 3 minutes or longer than 90 minutes. For the Bird and Lyft data, we are only able to infer unlinked trips; that is, we know the trip origins and destinations but do not know how they are linked. Therefore, we use data from all four operators when studying trip counts but use data for Spin and Lime only when examining trip characteristics such as duration and distance.

### 3.2 Docked Bikeshare Data

Data for the Capital Bikeshare system can be accessed freely through the following URL: https://www.capitalbikeshare.com/system-data. The trip data is available in a CSV format and the available data can be scrapped from real-time public APIs. Each trip's record includes the duration of the trip, start date and time, end date and time, start and end station, bike's ID number, and member type. We only kept trips with a duration of between 3 to 90 minutes, which resulted in a total of about 576,000 trips for analysis. Like shared e-scooters, we use the data scrapped at 6:00 am to analyze availability and accessibility.

### 3.3 Sociodemographic and Employment Data

We accessed demographic and employment data from the ACS and the Longitudinal Employer-Household Dynamics (LEHD) (https://lehd.ces.census.gov/) Database. The ACS is a demographics survey program conducted by the U.S. Census Bureau and provides a wide range of sociodemographic data including education, employment, income, population, and many other important statistics about the nation and its people. The main unit of analysis used in this study is block groups, which are generally defined to contain between 600 and 3,000 people. These data allow us to examine equity-related outcomes across block groups of different sociodemographic characteristics. Moreover, we collect the 2018 employment data at the block level from the LEHD Origin-Destination Employment Statistics and aggregate them to the block group level.

## 4. Analysis and Results

This section presents an equity analysis of the dockless e-scooter and station-based bikeshare services in Washington DC following the analytical framework discussed in Section 3. The results are summarized in Table 1 and Table 2. Much of the discussion below focuses on comparing the difference in mean values between block groups in different categories. To ensure that the comparisons are meaningful, we have applied two-sample t-test (the Welch's test) to examine statistical significance (results are presented in Appendix, Table A2). All following discussions are based on statistically significant results unless otherwise noted.

Table 1. E-scooter accessibility, availability, and idle time across block groups

|  | Availability | | | | | | Accessibility | | Idle Time (hours) | |
| --- | --- | --- | --- | --- | --- | --- | --- | --- | --- | --- |
|  | No. Daily Available E-scooters | | No. Daily Available E-scooters Per Resident | | No. Daily Available E-scooters Per Resident and Job | | Kernel Density | | | |
|  | Mean | Median | Mean ($\times 10^{-2}$) | Median ($\times 10^{-2}$) | Mean ($\times 10^{-2}$) | Median ($\times 10^{-2}$) | Mean | Median | Mean | Median |
| **Block groups divided by EEA status** | | | | | | | | | | |
| EEA | 23.68 | 9.46 | 1.66 | 0.62 | 0.97 | 0.56 | 114.79 | 40.13 | 5.19 | 3.9 |
| Non-EEA | 36.84 | 15.48 | 2.40 | 1.31 | 1.27 | 0.90 | 175.58 | 79.84 | 5.39 | 4.43 |
| **Block groups divided by median household income[a]** | | | | | | | | | | |
| Low | 20.18 | 6.26 | 1.52 | 0.40 | 0.95 | 0.37 | 78.65 | 26.76 | 5.34 | 4.69 |
| Middle | 37.11 | 14.70 | 2.36 | 1.20 | 1.18 | 0.86 | 183.71 | 85.65 | 5.27 | 4.09 |
| High | 26.03 | 14.70 | 1.83 | 1.26 | 1.15 | 0.87 | 132.66 | 68.46 | 5.47 | 4.36 |
| **Block groups divided by racial compositions[b]** | | | | | | | | | | |
| White | 43.59 | 6.26 | 2.72 | 1.50 | 1.21 | 1.00 | 203.48 | 126.35 | 5.33 | 4.36 |
| Black | 15.64 | 5.96 | 1.18 | 0.47 | 0.87 | 0.40 | 59.58 | 23.18 | 5.63 | 4.47 |
| No-Majority | 36.78 | 21.80 | 2.66 | 1.35 | 1.74 | 1.02 | 257.73 | 180.10 | 5.1 | 3.69 |
| Average | 30.10 | 13.35 | 2.02 | 0.95 | 1.12 | 0.71 | 144.51 | 59.52 | 5.35 | 4.33 |

Notes: a. Income level is defined by quartile, with the middle representing the middle 50% of median household incomes at block group level: low, ≤$49,222; middle, $49,223–$130,614; high, ≥$130,615.
b. Race with a dominant population (≥50%) within a block group.
c. An outlier is captured in the availability results at the block group level. The residents' population in a block group is very few while many available dockless e-scooters are deployed here to meet the travel demand of floating populations such as workers, which leads to a significantly higher number of daily available e-scooters per resident. This outlier raises the statistical results of corresponding block groups obviously and conceals most of the characteristics presented by the results, so we removed this outlier for dockless e-scooter services analysis.

Table 2. Bikeshare availability, accessibility and idle time across block groups

|  | Availability | | | Accessibility | | Idle Time (hours) | |
|---|---|---|---|---|---|---|---|
|  | No. Daily Available Bikes | No. Daily Available Bikes Per Resident | No. Daily Available Bikes Per Resident and Job | Kernel Density | | | |
|  | Mean | Mean (×10$^{-2}$) | Mean (×10$^{-2}$) | Mean | Median | Mean | Median |
| **Block groups divided by EEA status** | | | | | | | |
| EEA | 3.80 | 0.26 | 0.18 | 20.45 | 8.31 | 3.32 | 1.91 |
| Non-EEA | 6.16 | 0.41 | 0.23 | 29.33 | 14.46 | 2.56 | 1.27 |
| **Block groups divided by household income[a]** | | | | | | | |
| Low | 3.02 | 0.22 | 0.16 | 14.19 | 3.34 | 2.73 | 1.10 |
| Middle | 6.06 | 0.40 | 0.23 | 30.17 | 16.54 | 2.66 | 1.40 |
| High | 4.67 | 0.32 | 0.19 | 24.74 | 10.92 | 2.98 | 1.74 |
| **Block groups divided by racial composition[b]** | | | | | | | |
| White | 6.81 | 0.43 | 0.21 | 34.01 | 22.91 | 2.59 | 1.28 |
| Black | 2.99 | 0.22 | 0.18 | 12.66 | 2.91 | 4.00 | 2.44 |
| No Majority | 5.75 | 0.39 | 0.27 | 38.03 | 25.35 | 2.99 | 1.78 |
| Average | 4.95 | 0.33 | 0.20 | 24.79 | 10.76 | 2.74 | 1.42 |

Notes: See information provided in Table 1. The median values for all the bikeshare availability indicators are 0 because more than half of block groups in DC do not have bikeshare stations. Accordingly, we only present the mean values in this table.

**4.1 Equity in Service Availability**

We measure the availability of micromobility services at the block group level with three indicators: number of daily available vehicles, number of daily available vehicles per resident, and number of daily available vehicles per resident and job. The first indicator is a baseline measure and normalizing it by the resident population indicates the level of availability on a per resident basis. Considering that some visitors such as commuters are also micromobility users, we further use the residents plus jobs population to normalize the base indicator.

As the mean values of indicators shown in Table 1, block groups in non-EEA have more available dockless e-scooters than those in EEA. Middle-income block groups have the greatest number of daily available e-scooters, followed by the high-income block groups, and lastly the low-income block groups. The availability disparity between block groups classified by racial compositions is even larger: White-majority block groups have the highest number of daily available e-scooters, which is more than two times as many as that in Black-majority block groups. The results stay the same when we normalize the number of daily available e-scooters by the resident population. We further examine if results change when we normalize the number of daily available e-scooters by accounting for not only the resident population but also employment count (a proxy for activities and traveler flows). The major changes are with statistical significance. As shown in Table A2, the difference in daily available e-scooters per resident and job between block groups in different categories becomes statistically insignificant except for two comparisons: Black-majority block groups versus White-majority block groups and Black-majority block groups versus no-majority block groups. These results suggest that the disparities in e-scooter availability are most salient across block groups in different racial categories.

We also examine the median values of the availability indicators in case the results are biased by outliers. Similar results are obtained with some minor changes; specifically, no-majority block groups replace the White-majority block groups to rank top regarding the number of daily available e-scooters. Moreover, high-income block groups (replacing middle-income block groups) rank top in terms of the median values of daily available e-scooters per resident and daily available e-scooters per resident and job. These values are two or three times larger than those for low-income block groups. This finding is consistent with several previous studies of dockless micromobility systems which show that higher-income neighborhoods have greater micromobility availability (Mooney et al., 2019; Meng and Brown, 2021).

Regarding the bikeshare system, results in Table 2 suggest that fewer bikes are available in EEA block groups than in non-EEA block groups. Also, low-income block groups have fewer daily available bikes than middle-income block groups. This finding is consistent with a previous study of the bikeshare system in DC (Brown et al., 2019). We also find significant racial disparities in the availability of shared bikes across block groups: the number of daily available bikes in black-majority block groups is 3.0, less than half of that in White-majority block groups (6.8 bikes). The results remain the same after we normalize the number of daily available bikes by the resident population. These results indicate that fewer shared bikes are placed in low-income and black-majority neighborhoods and that residents of these neighborhoods are also disadvantaged in terms of bikeshare availability on a per capita basis. When further taking workers into consideration, we find that the disparities between advantaged and disadvantaged block groups are narrowed slightly; more importantly, the differences in the number of available e-scooters per resident and job between block groups in each category become statistically insignificant (see Table A2).

Compared to the station-based bikeshare system, the disparities in e-scooter availability between disadvantaged and advantaged neighborhoods are generally larger. We have measured these disparities with the ratio of e-scooter availability in each category of disadvantaged block groups and that in their respective counterparts. We do not present the results here due to space constraints.

In sum, we find that both micromobility systems have fewer available vehicles in block groups with an EEA status, low median household income, and a Black-majority population. The results are largely similar except for changes in statistical significance when the number of daily available vehicles is normalized by accounting for the resident population or by accounting for the resident population plus total employment.

### 4.2 Equity in Accessibility to Micromobility Options

The availability measures may have a modifiable areal unit problem (Su et al.,2011; Viegas et al.,2009; Wong, 2004), which is a type of statistical bias that occurs when point-based measures of spatial phenomena (e.g., location of available bikes or e-scooters) are aggregated into areal units (e.g., block groups). Specifically, the availability measures discussed above will underestimate the number of micromobility vehicles reachable from a neighborhood when some vehicles are located outside but close to the neighborhood boundary. To address this issue, we also evaluate supply with an accessibility indicator measured by kernel density.

Kernel density is a commonly applied method to measure accessibility to spatially distributed resources such as shared mobility services and food stores (Dai and Wang, 2011;

Yan et al., 2021). A higher kernel density estimation value indicates a greater level of accessibility. Here, the kernel density estimation works by fitting a smoothly curved surface over each bikeshare station or e-scooter, with the surface value being the highest at the station/scooter location (the kernel center) and decreasing with increased distance to the kernel center. The kernel density value eventually reaches zero when the distance to the kernel center reaches the predefined search radius. We follow Yan et al., (2021) to set the search radius at one-sixth of a mile for bikeshare stations and one-eighth of a mile for e-scooters, which are assumed as the distance thresholds for accessing these micromobility options. The output of kernel density estimation is a raster surface, and density at each output raster cell is calculated by accumulating the values of all the kernel surfaces where they overlay the raster cell center. After obtaining the kernel density estimation value, we further aggregate it to the block groups level by taking a zonal mean.

The spatial distribution of e-scooter and bikeshare accessibility is shown in Figure 4. E-scooter accessibility is highest in the central area of DC, and it gradually declines outwards toward the city boundary. Similar to the dockless e-scooter system, the bikeshare accessibility is highest in the central area of DC; it gradually decreases as we move away from the city center, but there are pockets of areas where bikeshare accessibility is higher than the surrounding areas.

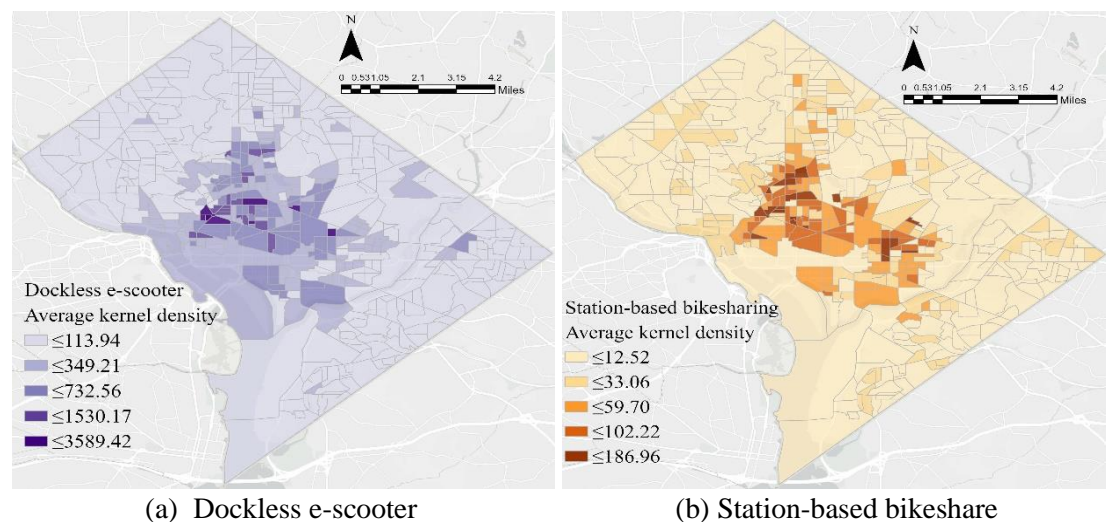

(a) Dockless e-scooter  (b) Station-based bikeshare
Figure 4. Spatial distribution of e-scooter (left) and bikeshare (right) accessibility.

We further examine how bikeshare and e-scooter accessibility vary across block groups divided by EEA status, median household income, and racial/ethnic compositions. Results are presented in Table 1 and Table 2. For the e-scooter system, the mean and median results of the indicator show a consistent pattern. Block groups not within EEA have higher kernel density values, which means higher levels of e-scooter accessibility than those in EEA. The level of e-scooter accessibility in middle-income block groups is the highest (183.71), followed by high-income block groups. Low-income block groups have the least access to e-scooter services, and their kernel density value (78.65) is less than half of that of the middle-income block groups. When considering block groups categorized by racial compositions, we find that black-majority block groups have the lowest accessibility, which is about a quarter of the average level of accessibility in no-majority block groups. We find no significant difference in e-scooter accessibility between White-majority and no-majority block groups. For the bikeshare system, we get similar results for except for one difference: while high-income block groups have lower bikeshare accessibility than middle-income block groups, (unlike the e-scooter system) the difference is statistically insignificant.

In general, these results are consistent with the results discussed above regarding the availability indicators. These results further confirm the results of some earlier studies that show a lower level of accessibility to micromobility systems in disadvantaged block groups such as those with an EEA status, a low median household income, and a Black-majority population (Brown et al., 2019; Mooney et al., 2019)

### 4.3 Equity in Micromobility Usage

We calculate three indicators to measure the usage pattern of micromobility services: number of trips, number of trips per resident, and number of trips per resident and job. Descriptive results are summarized in Table 3 to show the number of bikeshare and e-scooter trips across different types of block groups.

The results show that more e-scooter trips happen in non-EEA block groups than in EEA area. The middle-income and White-majority block groups, which have greater e-scooter availability and accessibility, generate more trips than block groups in their respective categories. There appear to be significant racial disparities in e-scooter use: 45% of the block groups in DC are Black-majority, but trips generated from these block groups only account for 5.4% of all e-scooter trips during the study period. When we normalize the number of trips by resident population, we get similar results. When we further refine the measure by accounting for total employment, the results show a slight change for block groups divided by household income. High-income block groups take over middle-income block groups to become the places that generate the most trips per resident and job on average. Given the right-skewed distribution of results, we also examine the median values. The results for the median values are largely the same as those for the mean values, except that the high-income block groups replace the middle-income block groups to become the category with the highest values for all three usage indicators. The number of e-scooters trips generated in high-income block groups is almost ten times that of the trips generated in low-income block groups.

The results for station-based bikeshare are largely similar to those for e-scooter services. During the study period, fewer bikeshare trips happened in EEA, low-income, and Black-majority block groups compared to their respective counterparts. We get similar trends when we normalize the number of bikeshare trips by accounting for the resident population or by accounting for residents plus workers. There is one major difference between the usage pattern of e-scooter services and that of bikeshare services. After normalized by resident population and total employment, the median/mean number of e-scooter trips is highest in high-income block groups whereas the median/mean number of bikeshare trips is highest in middle-income block groups. One may be tempted to interpret these results as suggesting that e-scooters serve a higher income population segment than station-based bikeshare does. However, this needs to be further verified by studies (e.g., survey research) that collect detailed socioeconomic information on micromobility users. Moreover, it should be noted that for both the e-scooter system and the bikeshare system, the differences in trips between high-income block groups and middle-income block groups are not statistically significant.

**Table 3.** Share e-scooter and bikeshare usage across block groups

| Mode | E-scooter[a] | | | | | | Bikeshare[e] | | |
|---|---|---|---|---|---|---|---|---|---|
| Trip Characteristics | No. Trips[b] | | No. Trips per Resident[c] | | No. Trips per Resident and Job[d] | | No. Trips | No. Trips Per Resident | No. Trips Per Resident and Job |
| | Median | Mean | Median | Mean | Median | Mean | Mean | Mean | Mean |
| **Block groups divided by EEA status** | | | | | | | | | |
| EEA | 57.00 | 609.10 | 0.04 | 0.43 | 0.03 | 0.16 | 452.70 | 0.31 | 0.15 |
| Non-EEA | 282.00 | 1269.20 | 0.21 | 0.79 | 0.17 | 0.30 | 1270.00 | 0.78 | 0.32 |
| **Block groups divided by household income** | | | | | | | | | |
| Low | 29.5 | 325.67 | 0.02 | 0.18 | 0.02 | 0.09 | 236.83 | 0.13 | 0.07 |
| Middle | 200.50 | 1196.30 | 0.16 | 0.78 | 0.12 | 0.26 | 1194.00 | 0.75 | 0.32 |
| High | 290.00 | 1005.00 | 0.22 | 0.69 | 0.17 | 0.29 | 781.10 | 0.53 | 0.23 |
| **Block groups divided by racial composition** | | | | | | | | | |
| White | 453.00 | 1820.00 | 0.35 | 1.17 | 0.28 | 0.38 | 1597.00 | 1.00 | 0.36 |
| Black | 32.00 | 145.83 | 0.03 | 0.09 | 0.02 | 0.07 | 159.05 | 0.10 | 0.09 |
| No majority | 434.00 | 712.10 | 0.21 | 0.55 | 0.17 | 0.28 | 796.40 | 0.57 | 0.34 |
| Average | 152.00 | 931.00 | 0.11 | 0.61 | 0.09 | 0.23 | 851.10 | 0.54 | 0.23 |

a. The data used to compute the e-scooter metrics are from Spin, Lime, Bird, and Lyft.
b. The sum of the number of e-scooter trips taken over the study period that begin and end at a block group.
c. The number of e-scooter trips taken over the study period per resident.
d. The number of e-scooter trips taken over the study period per resident and job.
e. The data source for station-based bikeshare trips is the Capital Bikeshare system.
Additional Notes: 1) The median values for bikeshare trips indicators are 0, since some block groups in DC do not have docking stations for bikes which means no bikeshare trips start or end here. Therefore, we only present the mean values in the table.
 2) An outlier is captured in the usage results at the block group level. The resident population in a block group is very few while it has a large number of floating populations such as workers, which generate a lot of trips in total. This leads to a significantly higher number of trips, which raises the statistical results of corresponding block groups obviously and conceals most of the characteristics presented by the results, so we removed this outlier for dockless e-scooter services analysis.

**4.4 Idle Time Analysis**

We calculate the idle time for each micromobility vehicle and then aggregate the results to the block groups level. The results are shown in the last columns of Table 1 and Table 2. E-scooters are generally idle for a shorter duration in EEA block groups than those in non-EEA block groups. This is result somewhat surprising, which we intend to explore further in a future study. The average idle time of e-scooters placed in low-income block groups is not significantly different from that of e-scooters placed in middle-income and high-income block groups, but e-scooters placed in middle-income block groups experience a shorter idle time than those placed in high-income block groups. Moreover, E-scooters in Black-majority block groups have a longer average idle time (5.63 hours) compared to those placed in White-majority (5.33 hours) and no-majority (5.1 hours) block groups.

We also calculate the median idle time of e-scooters in each category of block groups, and one change we observe is that low-income block groups replace high-income block groups to become places where e-scooters experienced the longest idle time. In other words, the utilization rate of e-scooters in low-income and Black-majority block groups is lower than that in their counterparts. Considered together with the findings on e-scooter availability/accessibility and usage, these results suggest that increasing supply alone is not likely effective in promoting e-scooter use in disadvantaged neighborhoods.

Regarding the bikeshare system, we find some differences between shared e-scooters and shared bikes regarding how their idle time varies across block groups. The average idle time of shared bikes in each category of block groups is nearly half of that of shared e-scooters. The idle time of shared bikes deployed within EEA block groups is shorter than that of shared bikes within non-EEA block groups, which is the opposite of the results for e-scooter. When block groups are categorized by household income, the differences in idle time across block groups are statistically significant. Share bikes have the shortest idle time in low-income block groups. On the other hand, consistent with the findings for shared e-scooters, shared bikes in Black-majority block groups have a longer idle time than those in White-majority or no-majority block groups; however, shared bikes in low-income block groups have a shorter median idle time than those in higher-income block groups. In other words, although shared bikes are less frequently used in minority neighborhoods, they have a relatively higher utilization rate in low-income neighborhoods. This is likely attributable to the "Capital Bikeshare for All" equity program, which allows qualified low-income individuals to use shared bikes for free (unlimited free rides for trips under 60 min) with a $5 annual membership fee. A further implication of these results is that the bikeshare equity program has not significantly benefited travelers in EEA or Black-majority neighborhoods.

**5. Discussions**
**5.1 E-scooter Services Increase Access to Shared Mobility Options in Disadvantaged Neighborhoods but Widen the Access Gap across Neighborhoods**

For a station-based bikeshare system, users need to access and return shared bikes at fixed docking stations. This means that neighborhoods without docking stations have little or no access to bikeshare service. Dockless e-scooters, due to their free-floating nature and their ability to be deployed in large quantities, can be easily expanded to these neighborhoods and enhance mobility for the residents. Also, unlike station-based bikeshare, e-scooters users can often avoid the "last mile" access problem because e-scooters can be parked at their destinations. Flexible and low-cost e-scooters allow them to penetrate wherever they are needed. Based on our analysis, we find that on average, e-scooter availability and accessibility are 2.7 times and 5.8 times that of bikeshare availability and accessibility, respectively, in D.C. neighborhoods. For instance, as shown in Table 1 and Table 2, the average accessibility of e-scooter services is about five times that of shared bikes in Black-majority block groups, which indicates that residents have better spatial access to e-scooters than station-based bikes.

In sum, we find that e-scooters have exaggerated the existing disparities in spatial access to micromobility services. EEA, low-income, and Black-majority block groups in general have a lower level of spatial accessibility to both e-scooter and bikeshare services. The arrival of e-scooters has widened these accessibility disparities.

**5.2 Compared to Bikeshare, Shared E-scooters Have a Higher Level of Spatial Accessibility but an Equivalent Number of Trips and Longer Idle Time**

Similar to the results for e-scooter supply (availability and accessibility), we find significant disparities in e-scooter use between disadvantaged and advantaged block groups. E-scooters are less used in EEA, low-income, and Black-majority block groups compared with their respective counterparts. Notably, although the average number of daily available e-scooters is over two times that of daily available bikes, the number of e-scooter trips is similar to that of bikeshare trips. This implies that on average, shared bikes are much more frequently used than shared e-scooters. The same conclusion can be reached by comparing the average idle time of shared e-scooters with that of shared bikes: the former is about twice that of the latter in most block groups. In block groups where the average idle time is particularly long, which indicates that the available e-scooters exceed the demand for e-scooter use, the operators should consider reducing the number of e-scooters placed there. A long idle time may lead to more cases of improper parking and vandalism activities of shared e-scooters.

The middle-income block groups have the highest accessibility to both micromobility systems among block groups of all income levels, where the mean idle time for shared e-scooters and bikes is also the lowest. This indicates that the users from middle-income block groups have a strong demand for micromobility services. This finding can help operators explore the expansion of service scope to other areas with potential markets for micromobility use. Accessibility to both e-scooter and bikeshare services in high-income block groups is high, but the average idle time of both types of micromobility vehicles is also long.

**5.3 Bikeshare's Equity Program Appears to be More Effective Than E-scooters' Equity Programs**

Since idle time is jointly shaped by both supply and demand, longer idle time may be caused by two factors: oversupply and lower demand. Theoretically, if we observe no significant difference in the supply of micromobility vehicles in two regions but the idle time of vehicles in region A is shorter than that in region B, it would indicate higher demand in region A. We observe such results when we compare low-income block groups and high-income block groups. Based on the results shown in Table A2, we find no significant difference in the number of daily available shared micromobility vehicles (e-scooters and bikes) between low-income and high-income block groups. Also, the average idle time for shared e-scooters in the two categories of block groups is not significantly different; however, the average idle time for bikes deployed in low-income block groups is significantly less than that in high-income block groups. The above results indicate that compared to high-income neighborhoods, low-income neighborhoods have a relatively higher demand for shared bikes but not for shared e-scooters. These differences are likely to result from the distinctive impacts of the two micromobility systems' equity programs. The equity program for bikeshare provides a much larger discount to low-income travelers compared to those offered by e-scooter companies. Qualified individuals can enjoy unlimited 60-min free bike rides at a $5 annual membership fee. By contrast, discounts offered by e-scooter companies range from a waiver of the $1 unlock fee to unlimited 30-min trips with a $5 membership fee per month. Moreover, e-scooter companies may have not promoted their equity programs among qualified individuals as much as Capital Bikeshare.[2] Within disadvantaged block groups, such as low-income, bikeshare operators have managed to increase the demand in low-income block groups and increase the usage frequency of bikes, but the equity program of dockless e-scooters doesn't help it gain higher appeal in low-income neighborhoods than high-income neighborhoods.

The averaged idle time of shared bikes in Black-majority block groups is considerably higher than that in non-Black-majority block groups. The same finding holds true for shared e-scooters, which on average have an idle time of 5.6 hours in Black-majority block groups. Considered together with the fact that micromobility accessibility and availability are lower in Black-majority block groups, these results suggest that the demand for micromobility use is weak. Therefore, it is quite likely that increasing vehicle supply alone would not lead to higher micromobility use; other coupling strategies such as promoting the enrollment of equity programs and reducing access barriers to micromobility services are also necessary.

**6. Conclusion**

This study conducts a comparative analysis of geographic equity in dockless e-scooters and station-based bike-share systems across neighborhoods in Washington DC. It addresses a major research need regarding the equity of shared e-scooter services. The dockless e-scooter system provides higher accessibility and has more daily available vehicles compared to station-based bikeshare. Our analysis shows that while e-scooter services improve access to micromobility options for underserved areas and low-income block groups, they widen the accessibility gaps between disadvantaged (EEA, low-income, and Black-majority) and advantaged neighborhoods. Unsurprisingly, we observe much more shared e-scooter trips and shared bike trips in advantaged neighborhoods than in disadvantaged neighborhoods. For instance, Black-majority block groups account for 45% of the block groups in DC, but only 6.2% bikeshare trips and 5.4% e-scooter trips in DC happened in these neighborhoods.

---

[2] Since 2020, District Department of Transportation (DDOT) has required all dockless micromobilty operators to offer free unlimited 30-min rides to qualified low-income travelers. However, the sign-ups for these low-income customer plans are small according to an informal conversation with DDOT staff.

Further analysis of idle time suggests that more shared bikes should be placed in low-income neighborhoods, considering that bikeshare accessibility is low but shared bikes in these neighborhoods have relatively short idle time compared to those in higher-income neighborhoods. This can be achieved by increasing the size of the bicycle fleet, moving some existing bikes from other areas to the target areas, or more rebalancing efforts to maintain the level of intended supply. However, increasing vehicle supply alone would probably not lead to much higher micromobility use in minority neighborhoods, especially for shared e-scooters. Policymakers should couple managing micromobility supply with other strategies. One strategy is to enhance the biking infrastructure to make travelers feel safer to use micromobility. Moreover, micromobility operators should accommodate the needs of individuals who are not tech-savvy or lack access to smartphones by lifting access restrictions to the service platforms. Finally, both public entities and micromobility operators should make efforts to reduce existing financial and physical barriers for underserved groups.

A major limitation of this research is that we lack information on the socio-economic attributes of micromobility users. For example, we cannot assure that micromobility trips starting from or ending at low-income block groups are made by low-income travelers. Future research should address this issue by integrating trip data (e.g., the GBFS data used here) with socioeconomic data that may be collected from survey questionnaires. Moreover, the current study has mainly focused on understanding whether and to what extent micromobility equity is an issue, and we have not thoroughly examined what factors cause inequality. Further work may explore the latter by applying more sophisticated statistical approaches such as regression analysis or Geographically Weighted Regression models (Aman et al., 2021; Meng and Brown, 2021). Finally, future research should consider directly engaging disadvantaged populations (e.g., through qualitative study approaches such as interviews and focus groups) to better understand their essential travel needs, their perceptions of and experiences with shared micromobility, the barriers that prevent them from using shared micromobility, and what resources and strategies can increase their likelihood to use shared micromobility options.

**Appendix**

We adopted the population-weighted approach from *(13)* to check if results change compared to the threshold approach used for identifying disadvantaged communities. In the population-weighted approach, outcomes are measured for each population group rather than groups of neighborhoods. Here, we first calculate the equity outcomes (e.g., availability, accessibility) for each population group across all block groups and then obtain a weighted mean using the sociodemographic data for each block group. Results are presented in Table A1.

Table A1. Availability and accessibility for groups with population weighting method

|  | Dockless e-scooter | | Station-based bikeshare | |
|---|---|---|---|---|
|  | Availability No. Daily Available E-scooters | Accessibility Kernel Density | Availability No. Daily Available Bikes | Accessibility Kernel Density |
| **Race** | | | | |
| White | 50.02 | 202.03 | 4.55 | 34.24 |
| Black | 25.48 | 93.87 | 2.82 | 17.05 |
| Asian | 58.10 | 242.61 | 4.83 | 37.08 |
| Other Race | 34.31 | 167.54 | 3.93 | 29.25 |
| **Household Income** | | | | |
| Low-income Households | 32.89 | 134.69 | 3.21 | 23.64 |
| Middle-income Households | 44.50 | 184.74 | 4.01 | 30.70 |
| High-income Households | 48.31 | 185.90 | 4.45 | 33.23 |

The results are largely similar to those obtained from the threshold approach. The Black population have the lowest level of accessibility to shared e-scooters and e-bike, nearly half of that enjoyed by the White population. Moreover, low-income households have lower accessibility to both shared micromobility options than middle-income and high-income households. Compared to results obtained from the threshold approach, one minor change is that the high-income households rather than the middle-income households have the highest accessibility to shared e-scooters and e-bikes.

In addition, to ensure that the comparison of differences in measured outcomes between block groups in different categories is meaningful, we applied the two-sample Welch's t-test to examine statistical significance. For the results of the test to be valid, following basic assumptions need to be met: observations must be independent; moreover, data in each group should be approximately normally distributed and obtained using a random sampling method. Empirically, two t-test is also valid for large samples from non-normal distributions. The data we used in this study are sufficient and each trip record is independent, thus the above assumptions are satisfied. Generally, the two samples should have approximately the same variance, but in this research, in case the variance of compared groups is not equal, we instead perform Welch's t-test which is used to compare the means between two independent groups when it is not assumed that the two groups have equal variances. From Table 1, Table 2 and Table 3, we could obtain the mean values of equity-related indicator, and the statistical test results are summarized in Table A2.

Table A2. Statistical test results for dockeless e-scooters and bikeshares across block groups.

| | Availability | | | | | |
|---|---|---|---|---|---|---|
| | No. Daily Available E-scooters | No. Daily Available E-scooters Per Resident | No. Daily Available E-scooters Per Resident and Job | Kernel Density | Idle Time | Usage[a] |
| **E-scooters** | | | | | | |
| EEA vs Non-EEA | **-2.59** | **-2.23** | -2.04 | **-3.29** | **-3.17** | **-2.59** |
| Low-income vs Middle-income | **-2.92** | **-1.96** | 1.11 | **5.71** | -1.18 | **-3.87** |
| Low-income vs High-income | -1.08 | -0.74 | -0.84 | **-3.11** | -1.74 | **-2.09** |
| Middle-income vs High-income | **1.97** | 1.51 | 0.23 | **2.17** | **-3.02** | -0.53 |
| Black-Majority vs White-Majority | **-5.09** | **-4.43** | **-2.34** | **-8.76** | **2.62** | **-6.01** |
| Black-Majority vs No-Majority | **-3.86** | **-2.73** | **-2.79** | **-5.69** | **3.59** | **-3.59** |
| White-Majority vs No-Majority | 0.95 | 0.11 | -1.75 | -0.91 | **2.29** | **3.48** |
| **Bikeshare** | | | | | | |
| EEA vs Non-EEA | **-2.93** | **-2.94** | -1.26 | **-3.65** | **28.36** | **-3.66** |
| Low-income vs Middle-income | **-3.57** | **-3.14** | -1.93 | **-6.35** | **2.60** | **-4.66** |
| Low-income vs High-income | -1.72 | -1.60 | -0.75 | **-4.16** | **-7.99** | **-2.32** |
| Middle-income vs High-income | 1.39 | 1.17 | 1.17 | 0.92 | **-12.92** | 1.44 |
| Black-Majority vs White-Majority | **-4.49** | **-4.02** | -1.09 | **-9.32** | **26.72** | **-5.92** |
| Black-Majority vs No-Majority | **-2.30** | **-2.01** | -1.49 | **-5.45** | **16.12** | **-3.41** |
| White-Majority vs No-Majority | 0.78 | 0.51 | -0.87 | 0.78 | **-10.72** | **2.65** |
| **E-scooters vs Bikeshare** | | | | | | |
| Within EEA | **7.32** | **5.80** | **7.35** | **8.74** | **31.51** | -1.00 |
| Within Non-EEA | **7.03** | **8.52** | **9.91** | **10.49** | **96.07** | 0.00 |
| Within Low-income | **4.29** | **3.77** | **4.13** | **5.56** | **50.25** | -0.60 |
| Within Middle-income | **7.26** | **7.52** | **10.18** | **10.60** | **70.64** | -0.01 |
| Within High-income | **5.66** | **6.16** | **6.88** | **7.16** | **43.12** | -0.60 |
| Within Black-Majority | **6.11** | **5.04** | **5.94** | **10.91** | **97.97** | 0.29 |
| Within White-Majority | **7.13** | **7.75** | **10.77** | **8.51** | **13.43** | -0.61 |
| Within No-Majority | **5.96** | **4.42** | **4.99** | **6.58** | **20.53** | 0.35 |

Note: a. Number of total trips.
b. Numbers in the table are t values, and numbers in bold are statistically significant at the 0.05 level.